\documentstyle[epsfig,preprint,aps]{revtex}
\begin{document}
\draft
\def\bge{\begin{equation}}
\def\ene{\end{equation}}
\def\bg{\begin{eqnarray}}
\def\en{\end{eqnarray}}
\def\nn{\nonumber}
\def\bsig{{\bar \sigma}}
\def\bomg{{\bar \omega}}
\def\amq{|{\vec q}|}
\title{Spectral change of $\sigma$ and $\omega$ mesons in a dense 
nuclear medium}
\author{K. Saito}
\address{Tohoku College of Pharmacy, Sendai 981-8558, Japan}
\author{K. Tsushima and A.W. Thomas}
\address{Department of Physics and Mathematical Physics\break
 and\break
Special Research Centre for the Subatomic Structure of Matter,\break
University of Adelaide, SA 5005, Australia}
\maketitle

\vspace{-9.7cm}
\hfill ADP-98-71/T338
\vspace{9.7cm}

\begin{abstract}
The spectra of the scalar ($\sigma$) and vector ($\omega$) mesons 
in nuclear matter are studied in detail using the 
quark-meson coupling (QMC) model. It is shown that  
above normal nuclear matter density the effects of $\sigma$-$\omega$ 
mixing and the decay of the $\sigma$ into $2\pi$ change  
the spectra of $\sigma$ and $\omega$ mesons considerably.  
As in Quantum Hadrodynamics 
we find a remarkable spectral change in the $\sigma$ 
meson and the longitudinal mode of the $\omega$ meson, namely 
a two-peak structure. 
\end{abstract}


Recently, medium modifications of hadron properties in a nucleus, 
which are in some sense precursors of the QCD phase transition, 
have received a deal of attention~\cite{qm97}.  
In particular, one of the most 
interesting phenomena is the spectral change of hadrons in a 
nuclear medium.  
At present lattice simulations have mainly been  
performed for finite temperature ($T$), with zero chemical 
potential~\cite{latt}.  
Therefore, many authors have investigated hadron properties at finite 
nuclear densities ($\rho_B$) using effective 
theories~\cite{qm97,mumodels,qmc1,qmc2,qcdsr}.  

In this paper, using one of these effective theories, namely 
the quark-meson coupling (QMC) model~\cite{qmc1,qmc2}, we will study the 
spectral change of the iso-scalar scalar ($\sigma$) and iso-scalar 
vector ($\omega$) mesons at finite $\rho_B$ (and $T=0$) in relativistic 
random phase approximation (RRPA).  
In Quantum hadrodynamics (QHD)~\cite{qhd} nuclear matter 
consists of {\em point-like\/}
nucleons interacting through the exchange of point-like $\sigma$
and $\omega$ mesons.  On the other hand, 
the QMC model could be viewed as an extension of QHD, where  
the mesons couple to confined quarks (not to point-like nucleons) and the
nucleon is described by the MIT bag model.  The QMC then yields an
effective Lagrangian for a nuclear system~\cite{qmc1,qmc2}, which has the 
same form as that in QHD with a {\em density dependent\/} coupling constant 
between the $\sigma$ and the nucleon (N) -- instead of a fixed value.  
Indeed, from the point of view of the energy of a nuclear system, the key 
difference between QHD and QMC lies in the $\sigma$-N coupling 
constant, $g_s$.  Although this difference may seem subtle, it leads to 
many attractive results~\cite{qmc1,qmc2}.  Therefore,   
it would clearly be very interesting to investigate the spectra of the 
mesons in a relativistic framework, including the structural changes 
of the nucleon in-medium, and compare the QMC results with those given by 
QHD~\cite{prp}.  

First, let us briefly summarize the calculation of the $\sigma$ and 
$\omega$ meson spectra in symmetric nuclear matter, using QHD~\cite{prp}.  
The starting point is the lowest order polarization insertion in the 
meson propagator.  It describes the coupling of the meson, of momentum 
$q$, to a particle-hole or nucleon-antinucleon excitation in a nuclear 
medium.  
The polarization insertions for the $\sigma$, $\Pi_s$, and the $\omega$, 
$\Pi_{\mu \nu}$, are respectively given as 
\bg
\Pi_s(q) &=& -ig_s^2 \int \frac{d^4k}{(2\pi)^4}\mbox{Tr}[G(k)G(k+q)]
     + \frac{3}{2} ig_{\sigma\pi}^2 m_\pi^2 \int
     \frac{d^4k}{(2\pi)^4} \Delta_\pi(k) \Delta_\pi(k+q), \label{pis} \\ 
\Pi_{\mu \nu}(q) &=& -ig_v^2 \int \frac{d^4k}{(2\pi)^4}
     \mbox{Tr}[G(k) \gamma_\mu G(k+q) \gamma_\nu], \label{piv}
\en
where we have added the contribution from the pion-loop to $\Pi_s$ (the
second term on the r.h.s. of eq.(\ref{pis})) in order to treat
the $\sigma$ more realistically, and 
$g_v$ and $g_{\sigma\pi}$ are, respectively, the 
the $\omega$-N and $\sigma$-$2\pi$ coupling constants.  
The pion propagator, $\Delta_\pi$, is given
by $1/(q_\mu^2 - m_\pi^2 + i\epsilon)$ with the pion mass,
$m_\pi$ (=138 MeV), and 
$G(k)$ is the self-consistent nucleon propagator (with momentum $k$)
in relativistic Hartree approximation (RHA) given as
\bg
G(k) &=& G_F(k) + G_D(k), \nn \\
   &=& (\gamma^\mu k^*_\mu + M^*) \left[
   \frac{1}{k^{*2}_\mu - M^{*2} + i\epsilon} + \frac{i\pi}{E^*_k}
   \delta(k^*_0 - E_k^*) \theta(k_F - |{\vec k}|) \right].
\label{propn}
\en
Here $k^{*\mu}=(k^0 - g_vV^0, {\vec k})$ ($V^0$ is the mean value of
the $\omega$ field), $E_k^*=\sqrt{{\vec k}^2 + M^{*2}}$ ($M^*$ is
the effective nucleon mass in matter) and $k_F$ is the Fermi momentum.

Using the nucleon propagator we can separate the polarization insertion
into two pieces: one is the density dependent part, $\Pi^D$, 
which involves at least one power of $G_D$, and the other is the
vacuum polarization insertion, $\Pi^F$, which involves only
$G_F$.  The former is finite, but the latter is divergent and  
must be renormalized.  We here choose to renormalize such that: 
for the $\omega$, $\Pi_{\mu \nu}^F$ vanishes at $q_\mu^2 = m_\omega^2$ 
and $M^* = M$, and for the $\sigma$, 
$\Pi_s^F = \frac{\partial}{\partial q_\mu^2} \Pi_s^F = 0$  
at $q_\mu^2 = m_\sigma^2$ and $M^* = M$, where 
$m_\omega$ (783 MeV), $m_\sigma$ (550 MeV) and $M$ (939 MeV) are 
respectively the free masses of the $\omega$, $\sigma$ and the nucleon.  
The explicit, analytic 
expressions for not only the density dependent parts, but also 
the vacuum polarization insertions, are given in Refs.\cite{prp,qes,lim}.   

At finite density the scalar and the time component 
of the vector channels are allowed to mix.   
This is described by the time component of 
the mixing polarization insertion: 
\bge
\Pi_m(q) = ig_sg_v \int \frac{d^4k}{(2\pi)^4}
     \mbox{Tr}[G(k) \gamma^0 G(k+q)].
\label{pim}
\ene
In this case there is no vacuum polarization and $\Pi_m$ 
vanishes at zero density.  (The explicit form can be also found in
Ref.~\cite{lim}.)

The RRPA for the meson propagators involves summing of the ring diagrams 
to all orders.  This summation has been discussed by many
authors~\cite{qhd,prp,qes,lim,chin}.  In this paper, using RRPA  
we calculate the spectral functions for the mesons, rather than searching 
for zeros of the inverse of the propagators, because the mesons 
(especially, the $\sigma$ meson) have large widths in nuclear 
matter~\cite{chiku}.  The spectral functions of 
the $\sigma$ meson, $S_s(q)$, and the transverse (T), $S_T(q)$, and 
longitudinal (L), $S_L(q)$, modes for the $\omega$ meson are 
respectively given by~\cite{prp} 
\bg
S_s(q) &=& - \frac{1}{\pi} {\Im}m\left[
\frac{\Delta_0(1-d_0\Pi_L)}{\epsilon_{SL}} \right], \label{spcs} \\
S_T(q) &=& - \frac{1}{\pi} {\Im}m\left[
\frac{d_0}{1 - d_0 \Pi_T} \right],   \label{spct} \\
S_L(q) &=& - \frac{1}{\pi} {\Im}m\left[
\frac{d_0(1-\Delta_0\Pi_s)}{\epsilon_{SL}} \right]. \label{spcl} 
\en
Here $\Pi_T = \Pi_{11}$ or $\Pi_{22}$, $\Pi_L = \Pi_{33}-\Pi_{00}$ 
(we choose the direction of ${\vec q}$ as the $z$-axis) and  
$\epsilon_{SL}$ is the longitudinal dielectric function 
\bge
\epsilon_{SL} = (1 - d_0 \Pi_L)(1 - \Delta_0 \Pi_s) -
(q_\mu^2/q^2) \Delta_0 d_0 \Pi_m^2, \label{long}
\ene
with $q=|{\vec q}|$, and the free meson propagators for the $\sigma$ and
$\omega$ mesons are 
\bge
\Delta_0(q) = \frac{1}{q_\mu^2 - m_\sigma^2 + i\epsilon}  \ \ \
\mbox{and} \ \ \
d_0(q) = \frac{1}{q_\mu^2 - m_\omega^2 + i\epsilon}.  
\label{freeo}
\ene
In Ref.~\cite{prp} the meson masses and the spectral functions in QHD were 
reported in detail within RRPA.  

Now we are in a position to include the effect of changes in the
internal structure of the nucleon in-medium.  
In order to do so, we consider the following
modifications to the approach just obtained for QHD:

\noindent(1) meson-nucleon vertex form factor

In QHD the interactions between the mesons and nucleon are {\em point-like}.
However, since both the mesons and nucleon are composite they have finite
size.  As the simplest example, one could
take a monopole form factor~\cite{bonn} at each meson-N vertex:
\bge
F_N(Q^2) = \frac{1}{1 - q_\mu^2/\Lambda_N^2} ,
\label{formn}
\ene 
with a cut off parameter $\Lambda_N = 1.5$ GeV.  In principle,
one could self-consistently calculate the form factor within QMC~\cite{ff}. 
However, as such changes are not expected to make a big difference, we
use eq.(\ref{formn}) in the following calculation.

\noindent(2) density dependence of the coupling constants

In QMC the confined quark in the nucleon couples to the $\sigma$ field
which gives rise to an attractive force.
As a result the quark becomes more
relativistic in a nuclear medium than in free space.  This implies that
the small component of the quark wave function, $\psi_q$, is enhanced in
medium~\cite{qmc1,qmc2}.  
The coupling between the $\sigma$ and nucleon is therefore expected to be
reduced at finite density because it is given in terms of the quark scalar
charge, $\int_{Bag} dV\/{\bar \psi}_q \psi_q$.  
On the other hand, the coupling between the vector meson and nucleon
remains constant, because it is related to the baryon number, which is
conserved. 

Next we consider the widths of the mesons.  
In eq.(\ref{pis}) we introduced the $\sigma$ decay into two pions.  For 
the pion-loop, we have added an appropriate counter term to the lagrangian, 
and prescribe the renormalization in free space.  We require the
condition~\cite{prp}: ${\Re}e\Pi_{2\pi}(q_\mu^2) = 0$ at $q_\mu^2 = 
m_\sigma^2$, where $\Pi_{2\pi}$ is the pion-loop contribution given in 
eq.(\ref{pis}).   Furthermore, we introduce a form factor at the 
$\sigma$-$2\pi$ vertex, which is again taken to have a monopole form: 
$F_\pi(q_\mu^2) = (\Lambda_\pi^2 - m_\sigma^2)/(\Lambda_\pi^2 - q_\mu^2)$ 
with $\Lambda_\pi$ = 1.5 GeV (we choose the same value as $\Lambda_N$, 
for simplicity).  
Since the $\sigma$ meson propagator in free space is given in terms 
of the pion-loop polarization insertion, the free width of the 
$\sigma$ meson is  
\bge
\Gamma_\sigma^0 = - \frac{{\Im}m\Pi_{2\pi}}{m_\sigma}
 \ \ \ \mbox{ at } \ q_\mu^2 = m_\sigma^2.  \label{swidth}
\ene
Choosing $\Gamma_\sigma^0 = 300$ MeV\footnote{Conventionally, 
the width may be larger than 300 MeV.  However, recent theoretical 
analyses suggest that it is near or below 300 MeV~\cite{chiku,swidth}.},  
we find $g_{\sigma\pi}$ = 18.33~\cite{prp}.  

It is known that while the $\omega$ has a narrow width in
free space, this should increase in nuclear matter~\cite{scatt}.
Therefore, it is interesting to study the effect of the
$\omega$ width on the meson spectra.  To see this we replace the
$i\epsilon$ term in $d_0$ in eq.(\ref{freeo}) by
$im_\omega\Gamma_\omega^*$ 
\bge
d_0 \to d_0 = \frac{1}{q_\mu^2 - m_\omega^2 +
      i m_\omega \Gamma_\omega^*}, \label{d0}
\ene
where
\bge
\Gamma_\omega^* = \Gamma_\omega^0 +
40 \mbox{(MeV)} \times (\rho_B/\rho_0),
\label{omgwdth}
\ene
is the in-medium $\omega$ width~\cite{scatt} (with the free width,
$\Gamma_\omega^0=9.8$ MeV). 

To study the meson spectral functions in nuclear matter, we first have
to solve the nuclear ground state within RHA.  
In QHD the total energy density for nuclear matter is written
as~\cite{prp,qes}
\bge
{\cal E} = {\cal E}_0 + \frac{1}{2\pi^2}M^2(M-M^*)^2 \left[
  \frac{m_\sigma^2}{4M^2} + \frac{3}{2} f(1-4M^2/m_\sigma^2)
-3 \right], \label{energy}
\ene
where ${\cal E}_0$ has the usual form (in RHA), given in Ref.~\cite{qhd}.
Here, $f(z) = 2\sqrt{-z} \tan^{-1}\frac{1}{\sqrt{-z}}$.  
Note that in Ref.~\cite{qhd} the renormalization condition on the nucleon
loops is imposed at $q_\mu^2$=0. The second term on the
r.h.s. of eq.(\ref{energy})~\cite{prp,qes} occurs because we chose
the renormalization condition for the $\sigma$ at $q_\mu^2=m_\sigma^2$.  
As measurable quantities cannot depend on this choice, our 
model gives the same physical quantities as those of Ref.~\cite{qhd}.

To take into account the modifications (1) and (2), we replace the
$\sigma$-N, $\omega$-N and $\sigma$-$2\pi$ coupling constants in 
eqs.(\ref{pis}), (\ref{piv}), (\ref{pim}) and (\ref{energy}) by
\bge
g_s \to g_s(\rho_B) \times F_N(q_\mu^2), \ \
g_v       \to g_v \times F_N(q_\mu^2) \ \mbox{ and } \
g_{\sigma\pi}   \to g_{\sigma\pi} \times F_\pi(q_\mu^2), \label{pff}
\ene
where the density dependence of $g_s(\rho_B)$ is given by solving the
nuclear matter problem self-consistently in RHA  
(see Ref.~\cite{qes}). 
As in QHD, we have two adjustable parameters in the present calculation:
$g_s(0)$ (the $\sigma$-N coupling constant at $\rho_B=0$) and $g_v$.

Requiring the usual saturation condition for nuclear matter,
namely ${\cal E}/\rho_B - M = - 15.7$ MeV at normal nuclear matter 
density ($\rho_0$ = 0.15 fm$^{-3}$),   
we determine the coupling constants $g_s^2(0)$ and $g_v^2$: 
$g_s^2(0)=61.85$ and $g_v^2=62.61$.  
In the calculation we fix the quark mass to be 5 MeV, 
while the bag parameters are chosen so as
to reproduce the free nucleon mass with the 
bag radius $R_0=0.8$ fm (i.e., $B^{1/4}=170.0$ MeV and
$z=3.295$~\cite{qmc1,qmc2}). 
This yields the effective nucleon mass $M^*/M=0.81$ at $\rho_0$
and the incompressibility $K=281$ MeV. 
(In the present work we do not consider the
possibility of medium modification of the meson properties 
at mean-field level~\cite{qmc2}.) 

\begin{figure}[htb]
\begin{center}
\epsfig{file=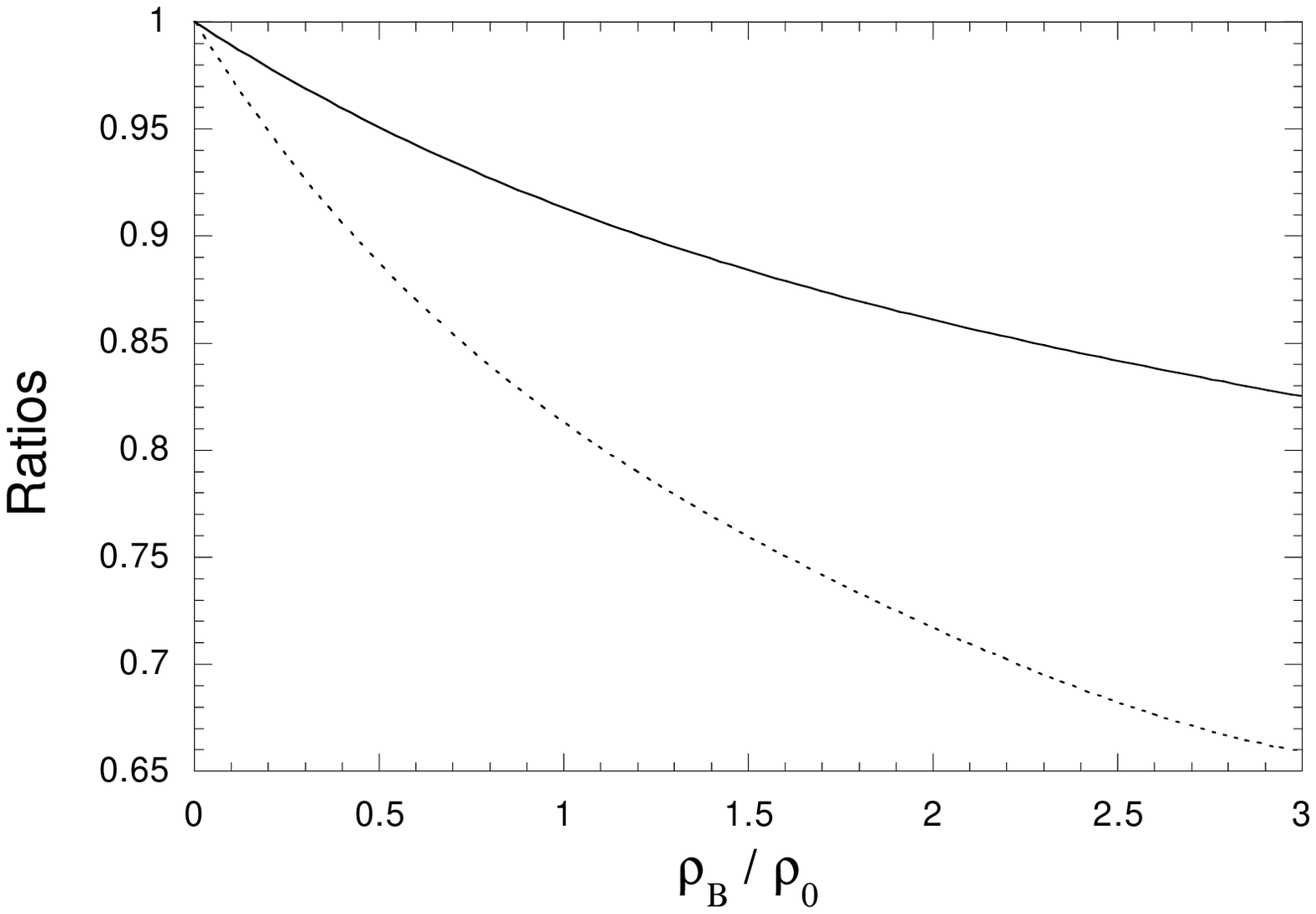,height=8cm}
\caption{Density dependence of $g_s(\rho_B)/g_s(0)$ and $M^*/M$.
The solid curve is for the ratio of the coupling constants, while
the dotted curve is for the ratio of the nucleon masses.
}
\label{f:ratios3}
\end{center}
\end{figure}

Now we present our main results.  First, in Fig.~\ref{f:ratios3}, we show
the density dependence of the coupling constant in QMC.   
At $\rho_0$, $g_s$ decreases by about 9\%.
The effective nucleon mass is also shown in the figure.

\begin{figure}
\begin{center}
\epsfig{file=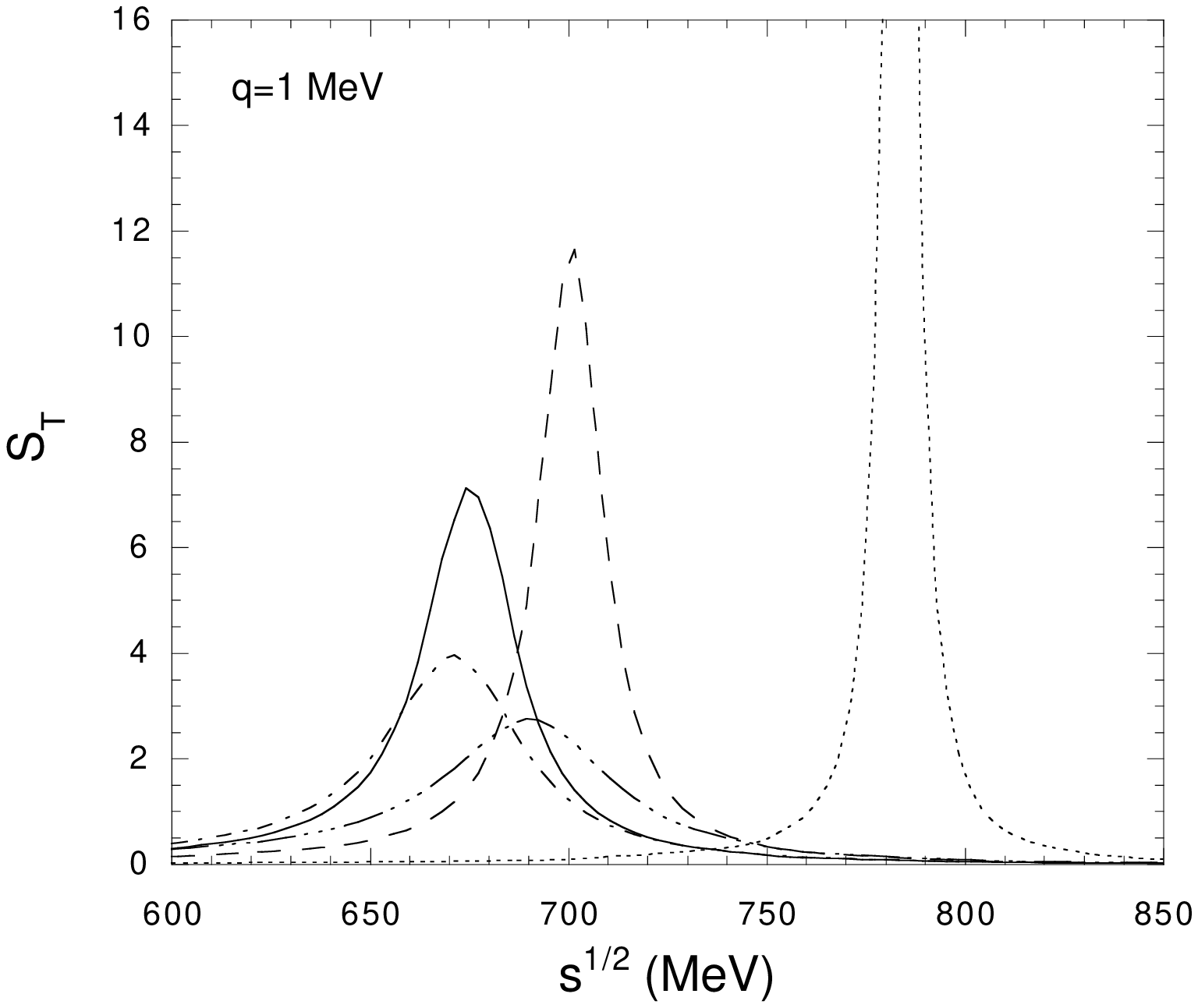,height=7cm}
\epsfig{file=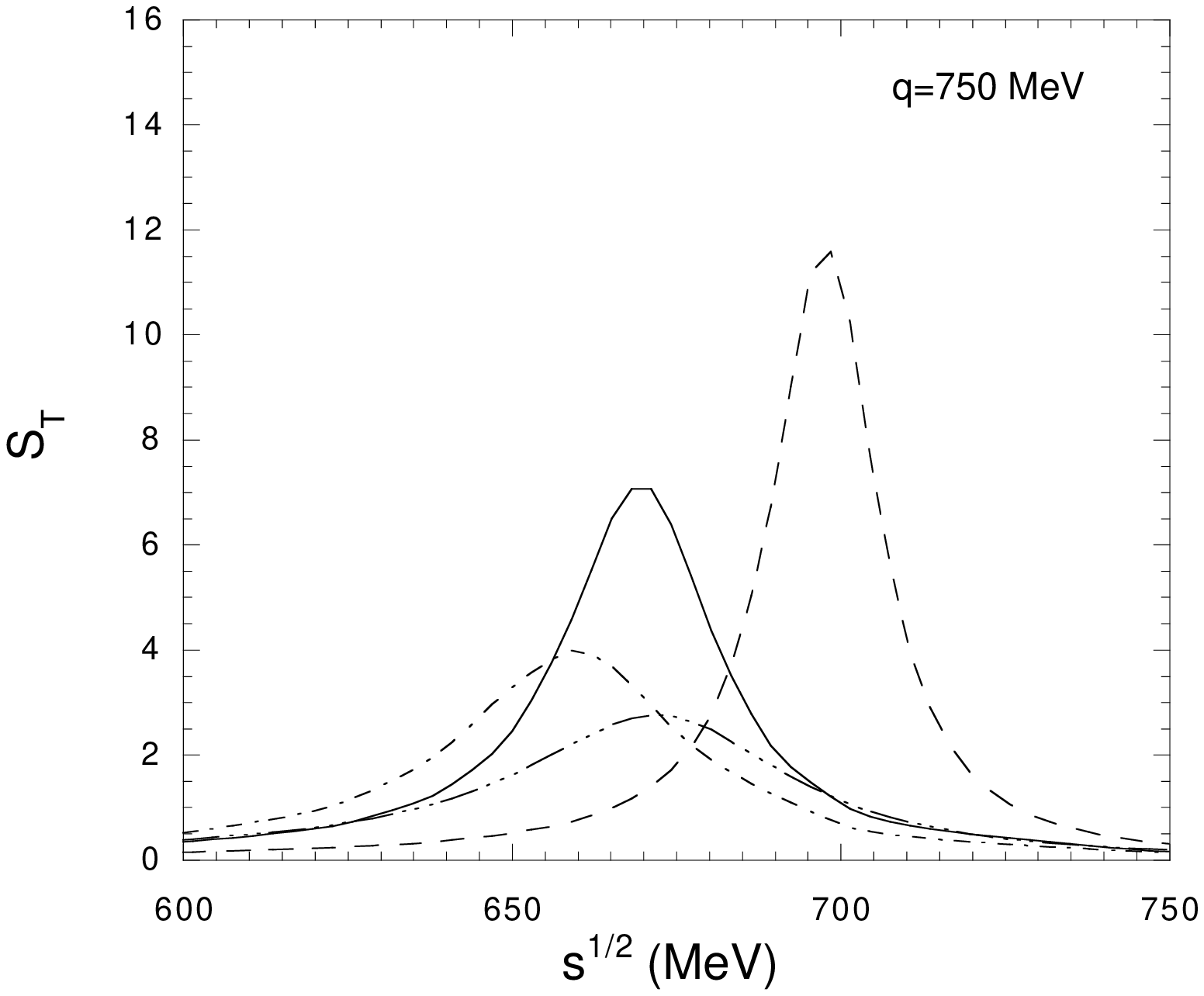,height=7cm}
\epsfig{file=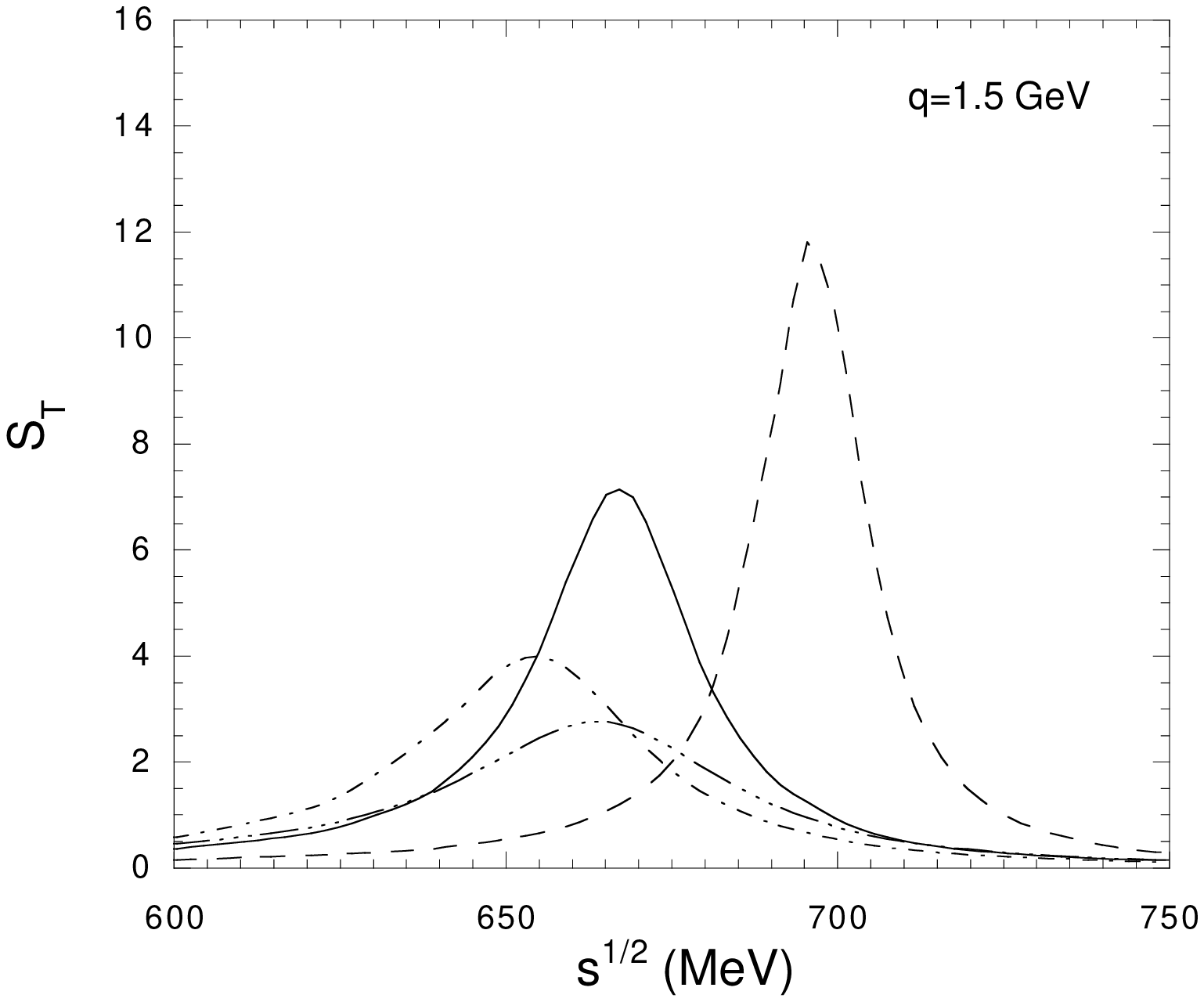,height=7cm}
\caption{
The transverse spectral
function, $S_T$, for the $\omega$ meson (in units of $M^{-2}$) at
three-momentum transfer $q$= 1, 750, 1500 MeV.
The dotted curve is for $\rho_B=0$, while 
the dashed (solid) [dot-dashed]
\protect\{3dot-dashed\protect\} curve is for $\rho_B/\rho_0$ = 0.5
(1) [2] \protect\{3\protect\}.
}
\label{f:ts}
\end{center}
\end{figure}

In Fig.~\ref{f:ts} we show the transverse spectral function of the 
$\omega$ meson.  For low three momentum transfer ($q$ = 1 MeV), we show 
the $\omega$ meson spectrum in free space (dotted curve), where 
the peak reaches about 37 (in units of $M^{-2}$) 
at the ``invariant'' mass 
$\sqrt{s} (= \sqrt{q_0^2 - |{\vec q}|^2}$) = 783 MeV.  
We see that the peak first 
moves downwards but comes back again to the high $\sqrt{s}$ 
side as the density increases.  The width of the T mode increases gradually, 
because of the in-medium $\omega$ width, $\Gamma_\omega^*$.  
The spectra for 
moderate ($q$ = 750 MeV) and high ($q$ = 1.5 GeV) three momentum transfer 
are very similar to each other. 

\begin{figure}
\begin{center}
\epsfig{file=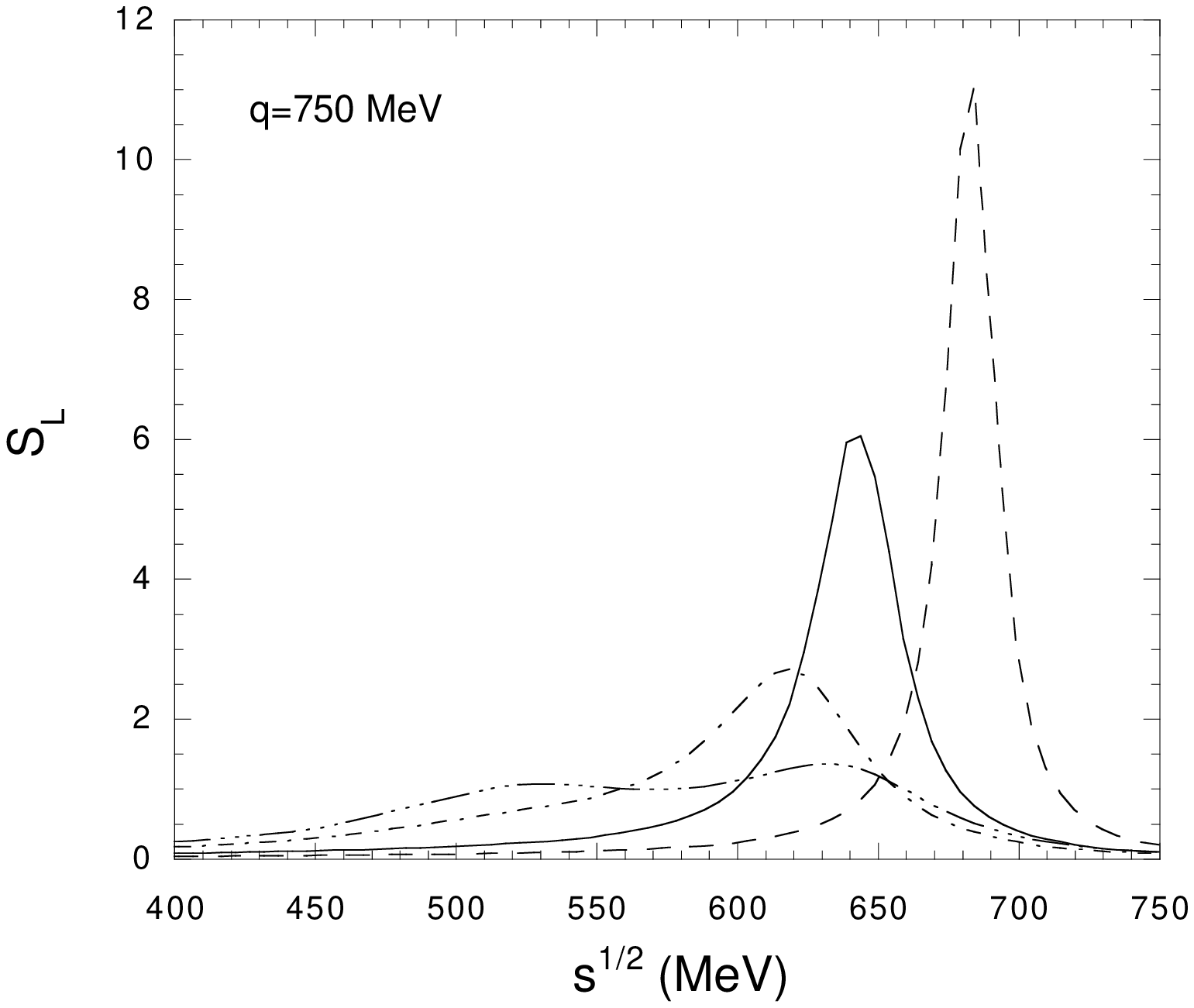,height=7cm}
\epsfig{file=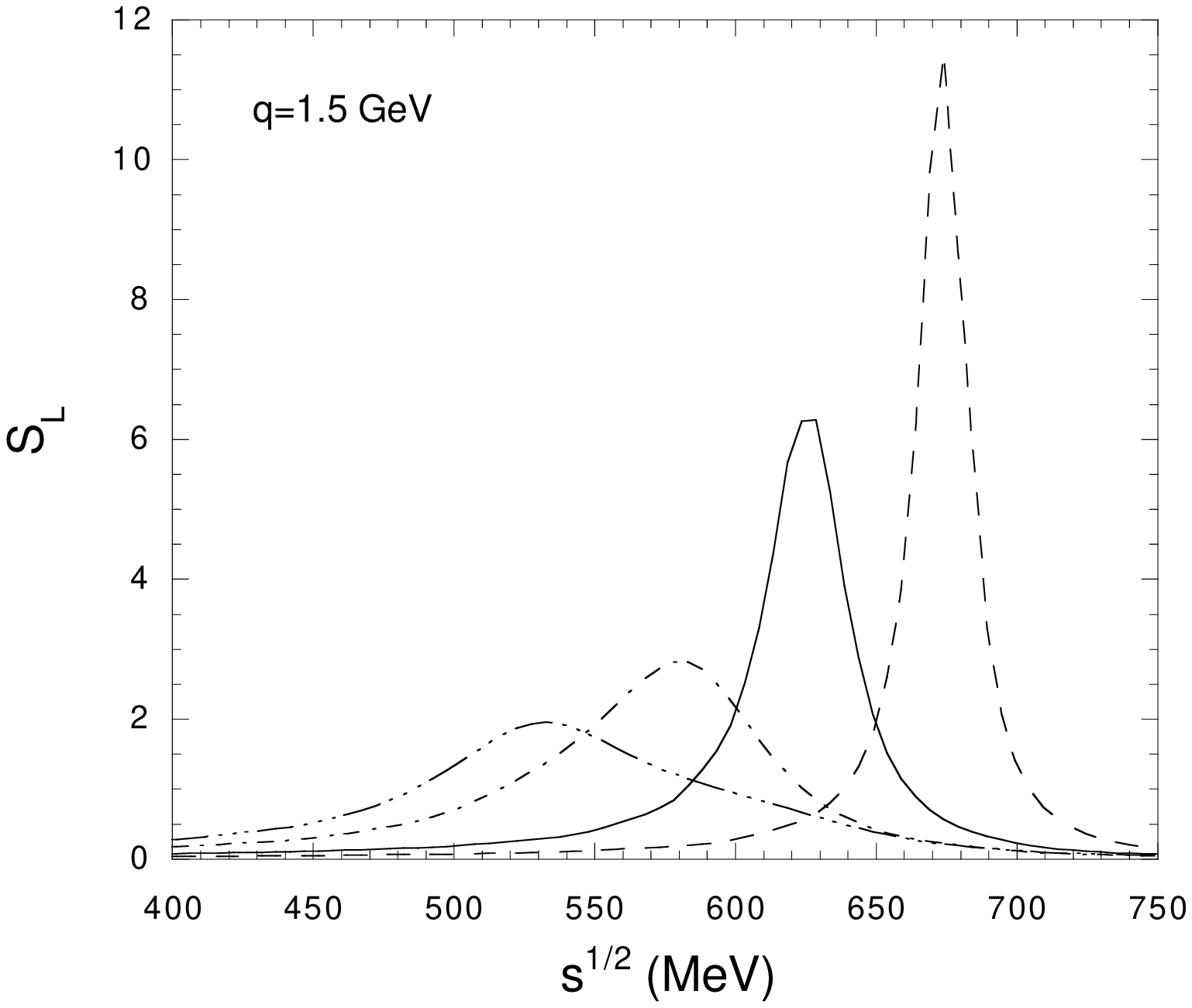,height=7cm}
\caption{
Same as Fig.~\protect\ref{f:ts} but for the longitudinal spectral
function, $S_L$, of the $\omega$ meson.  
}
\label{f:ls}
\end{center}
\end{figure}

\begin{figure}
\begin{center}
\epsfig{file=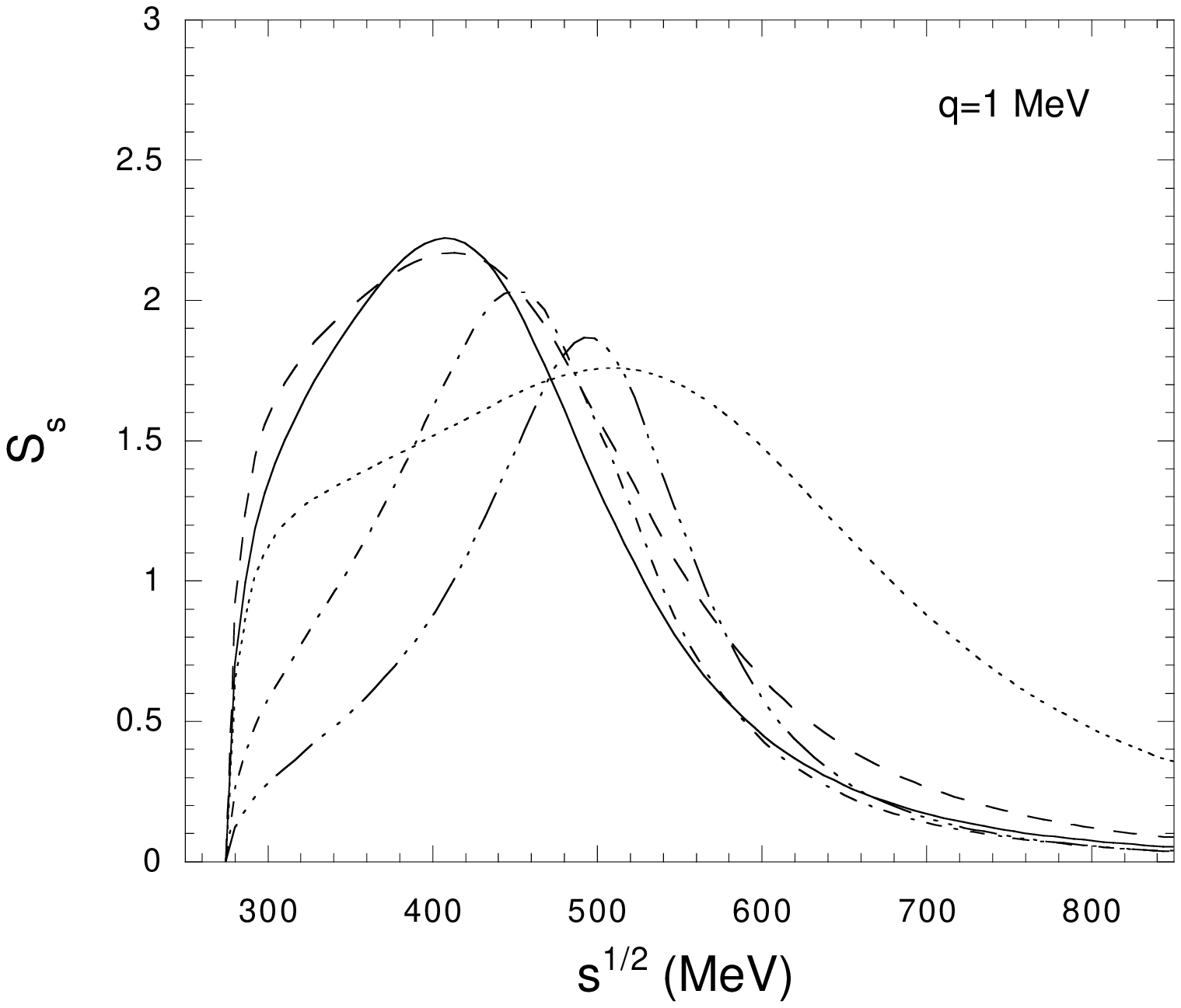,height=7cm}
\epsfig{file=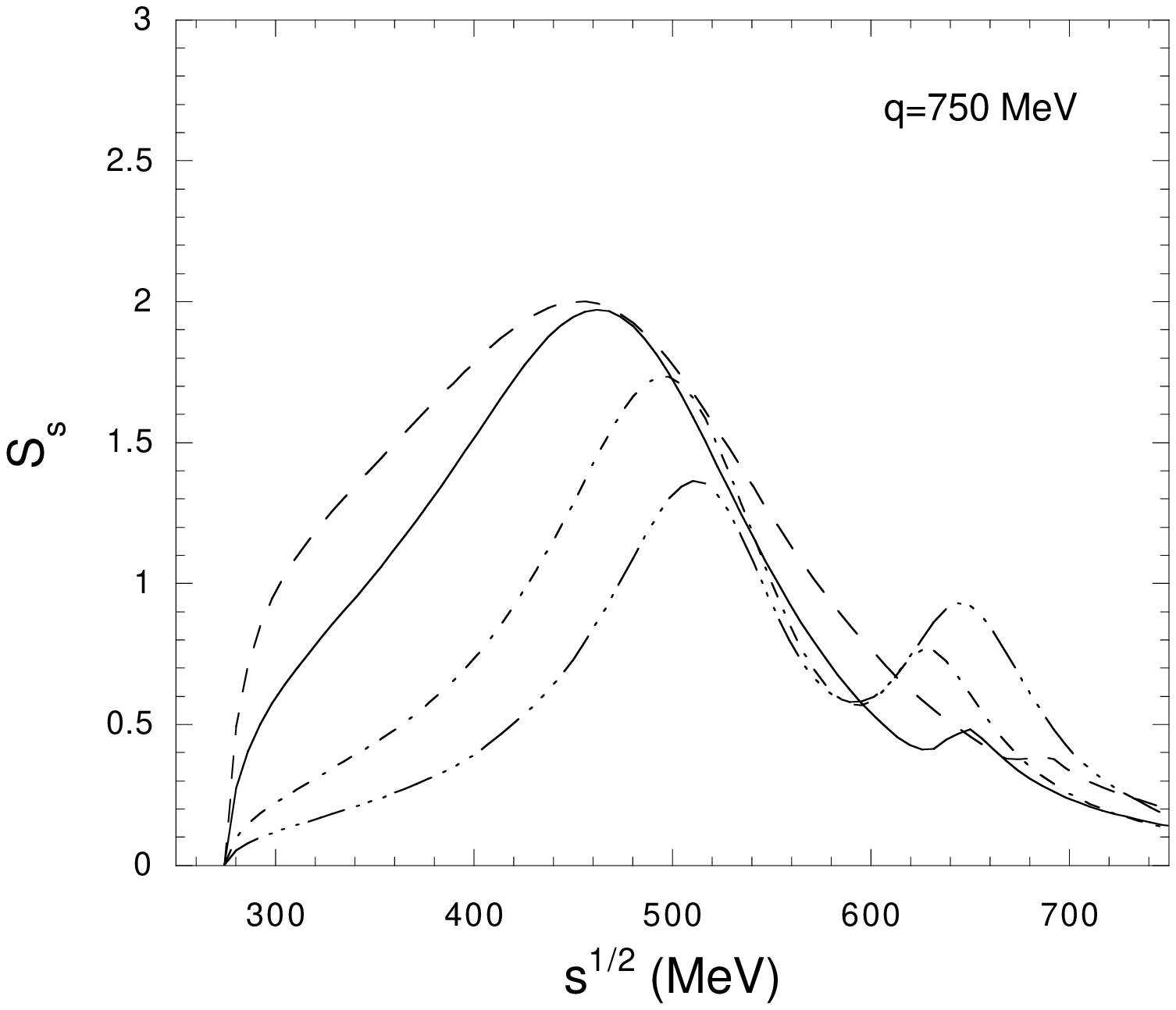,height=7cm}
\epsfig{file=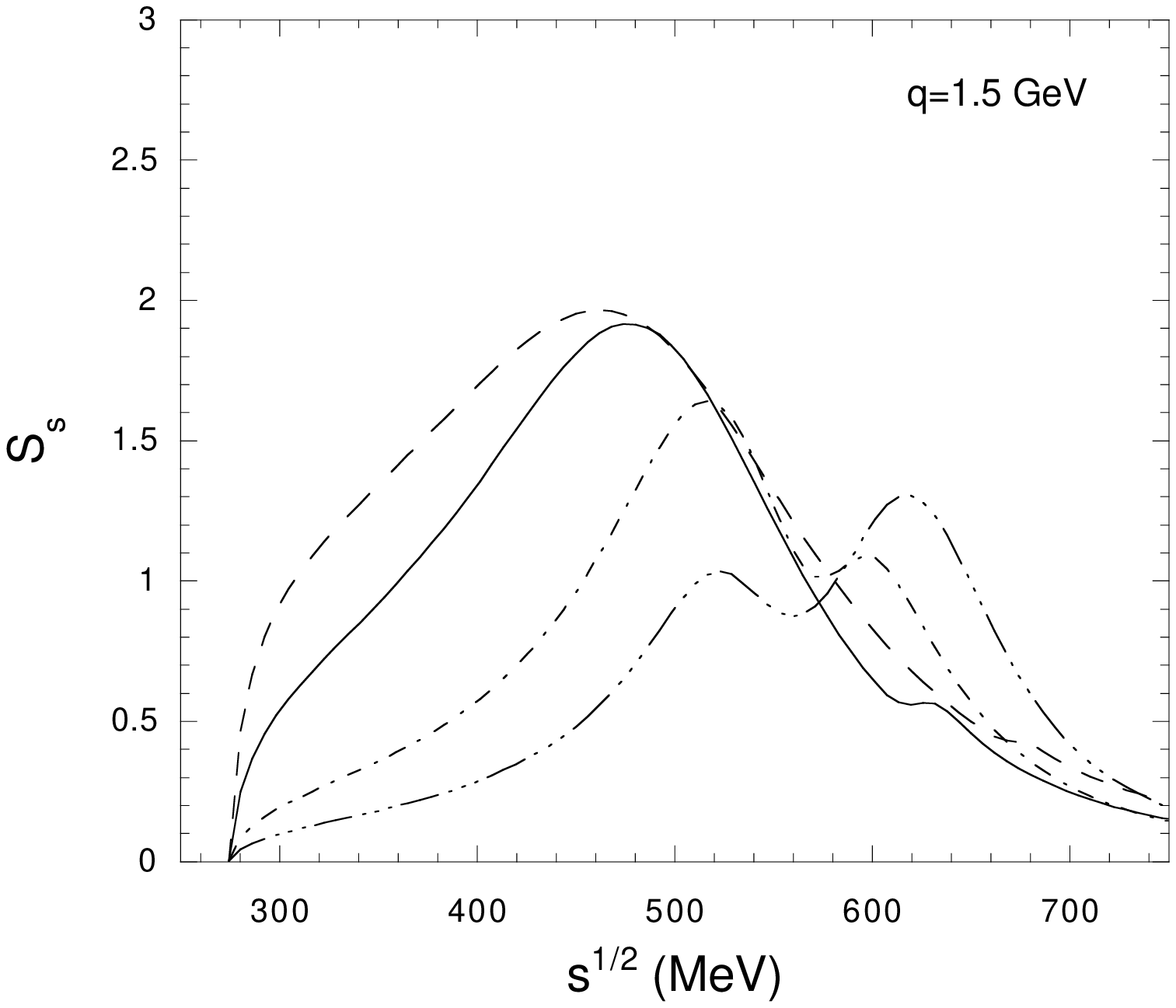,height=7cm}
\caption{
Same as Fig.~\protect\ref{f:ts} but for the spectral
function of the $\sigma$ meson, $S_s$.  
}
\label{f:ss}
\end{center}
\end{figure}

We present the longitudinal spectral function of the $\omega$ meson in 
Fig.~\ref{f:ls}.  The spectrum for low $q$ is almost the same as 
that in $S_T$, because there is no $\sigma$-$\omega$ mixing when $q$ 
vanishes.  The difference between $S_L$ and $S_T$ is caused by the 
$\sigma$-$\omega$ mixing at finite density.  It is very interesting 
that at moderate $q$ $S_L$ splits into two small, very broad peaks at 
high density, where the lower peak corresponds to the peak of the 
$\sigma$ channel (see Fig.~\ref{f:ss}).  
We can see that at high $q$ the higher peak 
almost disappears and only the lower peak remains at $\rho_B/\rho_0$ = 3. 

The spectral function of the $\sigma$ channel is shown in Fig.~\ref{f:ss}.  
The peak of the $\sigma$ meson first moves downwards and then 
moves back to the higher $\sqrt{s}$ side --- behaviour very similar 
to that seen for $S_T$.  The width of the $\sigma$ gradually shrinks at high 
density.  It is remarkable that at moderate or high $q$ the spectrum 
splits into two peaks at high density because of $\sigma$-$\omega$ mixing.  
The higher peak in $S_s$ corresponds to that in $S_L$, and it is enhanced 
at high $q$.  This two-peak structure has already been observed in 
the QHD calculation~\cite{prp}, where the separation into two peaks is 
seen more sharply than in the present calculation.  
This is because in QMC the increase of the $\omega$ width and 
the density dependence of the coupling constant moderate the 
$\sigma$-$\omega$ mixing in nuclear matter. 
 
To summarize, using QMC
we have calculated the spectral functions of the $\sigma$ and $\omega$ 
mesons in a dense nuclear medium.  
We have seen that the $\sigma$-$\omega$ mixing is quite 
important in dense matter (above $\sim \rho_0$), leading to an interesting 
spectral change in the $\sigma$ channel and the L mode of the $\omega$ 
channel --- namely a two-peak structure.   
Although the width of the $\sigma$ at finite density is smaller than 
that in free space, it is, still broad, and the 
amplitude of the spectrum itself is very small.  For the $\omega$ the 
spectrum becomes very broad as the density increases, and the amplitude is 
sensitive to the width of the $\omega$ in matter. 
We have already reported this remarkable structure in QHD~\cite{prp}, but  
in the present calculation this is moderated by the in-medium  
$\omega$ width and the decrease of the coupling  
constant in matter.  In future experiments (like 
dilepton production in heavy ion collisions~\cite{qm97}) 
it may be possible to obtain information on the spectral change of 
the mesons. While some experimental possibilities have 
been proposed for studying meson properties in hot/dense nuclear 
matter~\cite{chiku,kuni}, 
in particular, a spectral enhancement of the $\sigma$ channel 
near $2m_\pi$ threshold is studied in Ref.~\cite{kuni}, it may not be 
so easy to detect a clear signal of the change in a (cold) dense 
nuclear medium. 
Nevertheless, these are important challenge for the future.

\vspace{1cm}
This work was supported by the Australian Research Council and 
the Japan Society for the Promotion of Science. 
\clearpage
%
%
 
%
%
\newpage
\newpage
\newpage
\end{document}